\documentclass[twocolumn,pra,amsmath,amssymb,superscriptaddress]{revtex4-1}
\usepackage[utf8]{inputenc}
\usepackage{amsfonts,amsmath,amssymb,mathtools}
\usepackage{graphicx}
\usepackage{bm} 
\usepackage{mathrsfs}
\usepackage{subfigure}
\usepackage{textcomp}
\usepackage{microtype}
\usepackage{hyperref}
\usepackage[flushleft]{threeparttable}
\usepackage[per-mode=symbol]{siunitx}
\usepackage[version=3]{mhchem}

\usepackage{booktabs}
\DeclareSIUnit\intensity{\watt\per\centi\meter\squared}
\DeclareSIUnit\fieldstrength{\volt\per\centi\meter}
\usepackage{xspace}

\DeclareUnicodeCharacter{2009}{\,}

\newcommand{\degree}{\ensuremath{^\circ}}%

\newcommand{\Akplus}{\textup{Ak}^\textup{+}}
\newcommand{\Akmplus}{\textup{Ak}^{\prime +}}

\newlength{\figwidth}
\newlength{\figwidthwide}
\let\orgautoref\autoref
\providecommand{\Autoref}{%
  \def\equationautorefname{Equation}%
  \def\figureautorefname{Figure}%
  \def\subfigureautorefname{Figure}%
  \def\tableautorefname{Table}
  \def\sectionautorefname{Section}%
  \orgautoref}
\renewcommand{\autoref}{%
  \def\equationautorefname{Eq.}%
  \def\figureautorefname{Fig.}%
  \def\subfigureautorefname{Fig.}%
  \def\sectionautorefname{Sec.}%
  \orgautoref}

\usepackage{color}
\usepackage[normalem]{ulem}
\definecolor{darkgreen}{rgb}{0.0,0.7,0.0}

\begin{document}


\title{Laser-induced Coulomb explosion of heteronuclear alkali dimers on helium nanodroplets} 



\author{Simon H. Albrechtsen}
\thanks{These authors contributed equally to the work.}
\affiliation{Department of Physics and Astronomy, Aarhus University, Ny Munkegade 120, DK-8000 Aarhus C, Denmark}

\author{Jeppe K. Christensen}
\thanks{These authors contributed equally to the work.}
\affiliation{Department of Chemistry, Aarhus University, Langelandsgade 140, DK-8000 Aarhus C, Denmark}

\author{Rico Mayro P. Tanyag}
\affiliation{Department of Chemistry, Aarhus University, Langelandsgade 140, DK-8000 Aarhus C, Denmark}

\author{Henrik H. Kristensen}
\affiliation{Department of Physics and Astronomy, Aarhus University, Ny Munkegade 120, DK-8000 Aarhus C, Denmark}

\author{Henrik Stapelfeldt}
\email[]{henriks@chem.au.dk}
\affiliation{Department of Chemistry, Aarhus University, Langelandsgade 140, DK-8000 Aarhus C, Denmark}

\date{\today}

\begin{abstract}

A sample mixture of alkali homonuclear dimers, Ak$_2$ and Ak$^{\prime}_2$ and heteronuclear dimers, AkAk$^{\prime}$, residing on the surface of helium nanodroplets are Coulomb exploded into pairs of atomic alkali cations, ($\Akplus$,$\Akplus$), ($\Akmplus$,$\Akmplus$), ($\Akplus$,$\Akmplus$), following double ionization induced by an intense 50 fs laser pulse. The measured kinetic energy distribution $P(E_{\text{kin}})$ of both the $\Akplus$ and the $\Akmplus$ fragment ions contains overlapping peaks due to contributions from Coulomb explosion of the homonuclear and the heteronuclear dimers. Using a coincident filtering method based on the momentum division between the two fragment ions in each Coulomb explosion event, we demonstrate that the individual $P(E_{\text{kin}})$ pertaining to the ions from either the heteronuclear or from the homonuclear dimers can be retrieved, for both the $\Akplus$ and for the $\Akmplus$ fragment ions. This filtering method works through the concurrent detection of two-dimensional velocity images of the $\Akplus$ and the $\Akmplus$ ions implemented through the combination of a velocity map imaging spectrometer and a TPX3CAM detector. The key finding is that $P(E_{\text{kin}})$ for heteronuclear alkali dimers can be distinguished despite the simultaneous presence of homonuclear dimers. From $P(E_{\text{kin}})$ we determine the distribution of internuclear distances $P(R)$ via the Coulomb explosion imaging principle. We report results for \ce{LiK} and for \ce{NaK} but our method should should also work for other heteronuclear dimers and for differentiating between different isotopologues of homonuclear dimers.

\end{abstract}

\maketitle 

\section{Introduction and background}\label{sec:intro}

Nanometer-sized droplets of liquid helium provide a unique medium for studying structure and dynamics of atoms, molecules and clusters~\cite{stienkemeier_electronic_2001,choi_infrared_2006,yang_helium_2012,thaler_long-lived_2020,albertini_chemistry_2022}. While most of such dopants are located in the interior of the droplets, alkali atoms and small clusters thereof take a special role by residing on the droplet surface~\cite{dalfovo_atomic_1994,ancilotto_sodium_1995,toennies_superfluid_2004,barranco_helium_2006,stienkemeier_spectroscopy_2006,bovino_spin-driven_2009,guillon_theoretical_2011}. Notably alkali dimers and trimers have been extensively investigated for more than 25 years. Experimentally, the main methods used was absorption spectroscopy in which pulsed or continuous wave laser beams excited the dimers or trimers electronically with vibrational state resolution. It was such experimental studies that established that the dimers and trimers reside on the surface and, furthermore, that the dimers are mainly formed in the lowest-lying triplet state and the trimers in the lowest-lying quartet state~\cite{stienkemeier_laser_1995,higgins_spin_1996,higgins_photoinduced_1996,bruhl_triplet_2001,tiggesbaumker_formation_2007,mudrich_photoionisaton_2014,lackner_spectroscopy_2013}.

\begin{figure}
\includegraphics[width = 8.6 cm]{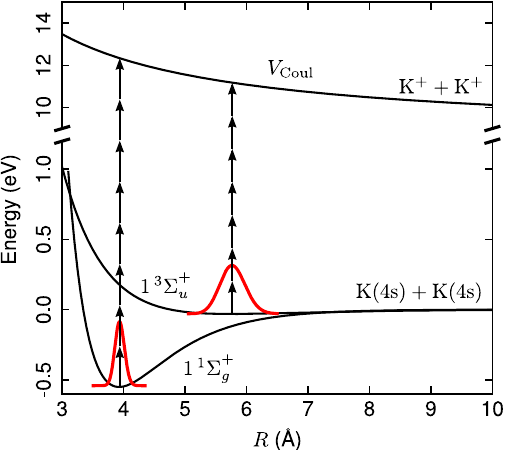}
\caption{Illustration of laser-induced Coulomb explosion of an alkali dimer using \ce{K2} as an example. The energy diagram depicts the potential curves for the 1~$^1\Sigma_{g}^+$ state~\cite{magnier_theoretical_2004} and the 1~$^3\Sigma_{u}^+$ state~\cite{bauer_accurate_2019} with the square of the corresponding vibrational ground state wave functions given in red. The potential curve for \ce{K2^{2+}} is shown as a Coulomb potential, see \autoref{eq:dication-Coulomb}. The black arrows represent the photons (not to scale) of the probe laser pulse to illustrate the multiphoton absorption process that causes double ionization of \ce{K2} and thereby Coulomb explosion into a pair of \ce{K+} ions.  \label{fig:potential-sketch}}
\end{figure}

Recently, it was demonstrated that Coulomb explosion induced by irradiation with an intense femtosecond laser pulse provides an alternative way to explore alkali dimers and trimers, for instance \ce{Na2} and \ce{Na3}, made of a single alkali species~\cite{kristensen_quantum-state-sensitive_2022,kristensen_laser-induced_2023}. The idea of the method in the case of an alkali dimer, the main topic of this paper, is the following, see also \autoref{fig:potential-sketch}. First, the dimer is doubly ionized through multiphoton absorption from an intense fs laser pulse typically with a central wavelength around 800 nm. Second, the resulting dication \ce{Ak2^{2+}} breaks apart into two \ce{Ak+} fragment ions:
\begin{equation}
\label{eq:Coulomb-idea}
\begin{aligned}
\ce{Ak2} + N\hbar\omega~\rightarrow~\ce{Ak2^{2+}}~+2e^-\\
\ce{Ak2^{2+}}~\rightarrow~\ce{Ak+}~+~\ce{Ak+}.
\end{aligned}
\end{equation}
Since the double ionization process removes the two valence electrons, \ce{Ak2^{2+}} has a closed-shell structure and therefore it is formed in only one state. As such, there is only a single, repulsive potential curve $V_\text{dicat}$ for \ce{Ak2^{2+}}. For the equilibrium internuclear distances of \ce{Ak2} in either the  1~$^1\Sigma_{g}^+$ or in the 1~$^3\Sigma_{u}^+$ state, $V_\text{dicat}$ is, to a good approximation, given by a Coulomb potential:
\begin{equation}
\label{eq:dication-Coulomb}
\begin{aligned}
V_\text{dicat} \approx V_\text{Coul} = \dfrac{14.4~\text{eV}}{R\text{(\AA)}}~+~2I_\text{p}(\text{Ak}),
\end{aligned}
\end{equation}
see \autoref{fig:potential-sketch}, where $I_\text{p}$(Ak) is the ionization potential of \ce{Ak}.

The single potential curve ensures a one-to-one correspondence between the initial internuclear distance $R$ of the dimer and the final kinetic energy $E_\text{kin}$ of each of the \ce{Ak+} fragment ions. If the dimer is made up of two alkali atoms with the same mass, for instance \ce{^{39}K2}, then energy and momentum conservation leads to $E_\text{kin} = \frac{7.2~\text{eV}}{R(\text{\AA})}$  in the Coulomb potential approximation.  One useful consequence of the one-to-one ($R$, $E_\text{kin}$) correspondence is that measurement of $E_\text{kin}$ makes it possible to distinguish between \ce{Ak2} in the 1~$^1\Sigma_{g}^+$ and in the 1~$^3\Sigma_{u}^+$ state~\cite{kristensen_quantum-state-sensitive_2022}. It was found that dimers are preferentially formed in the 1~$^3\Sigma_{u}^+$ state. The quantum state sensitivity was exploited to record laser-induced nonadiabatic alignment dynamics of alkali homonuclear dimers in both the triplet and in the singlet state~\cite{kranabetter_nonadiabatic_2023}.  With this method the distribution of internuclear distances $P(R)$ for the dimers in the two quantum states can also be determined~\cite{kristensen_laser-induced_2023}. This approach has been extended to fs time-resolved measurements of $P(R)$ for vibrating dimers~\cite{jyde_vibration_2024}.

In the current work, we explore Coulomb explosion of heteronuclear alkali dimers, AkAk$^{\prime}$, on the surface of helium droplets, where Ak and Ak$^{\prime}$ are two different alkali atoms, for instance Li and K. At first, it may seem like a straightforward extension of the former work on homonuclear dimers but due to the way the heteronuclear dimers are created, this is not the case. In practice, an AkAk$^{\prime}$ dimer is formed by passing He droplets through two serially arranged pickup cells. A droplet can then pick up an Ak atom in the first cell and an Ak$^{\prime}$ atom in the second cell, which may lead to the formation of a AkAk$^{\prime}$ heteronuclear dimer. The statistical nature of the pickup process implies, however, that some droplets pick up two Ak atoms or two Ak$^{\prime}$ atoms leading to formation of an \ce{Ak2} or an Ak$^{\prime}_2$ homonuclear dimer~\cite{higgins_helium_1998,mudrich_formation_2004}.

For the Coulomb explosion technique the complication of a mixed sample is that both the $\Akplus$ and the $\Akmplus$ fragment ions contain contributions from Coulomb explosion of the homonuclear dimers as well as of the heteronuclear dimer. The fragment ions originating from the homonuclear dimers and from the heteronuclear dimer will, in most cases, have overlapping kinetic energy distributions and, as a result, the one-to-one correspondence between $E_\text{kin}$ and $R$ is lost. Thus, the directly measured kinetic energy distribution may no longer be useful for identifying the quantum states of the dimer or for retrieving $P(R)$. The purpose of this paper is to show that it is possible to extract the kinetic energy distribution $P(E_{\text{kin}})$ of the fragment ions for both the homonuclear dimers and for the heteronuclear dimer. Hereby it becomes possible to determine $P(R)$ for the heteronuclear dimer.

\section{Experimental Setup}\label{sec:setup}

\begin{figure*}
\includegraphics[width = 17.8 cm]{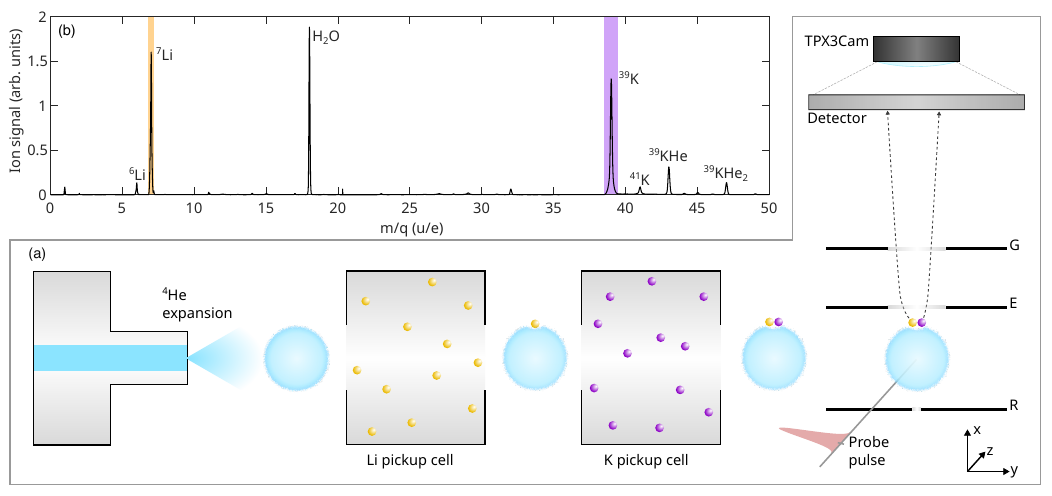}
\caption{(a) Schematic of the doping process and following probing process in the experiment (see the text). Not to scale. (b) Mass-to-charge spectrum obtained when the probe pulse irradiates droplets that passed through the two pickup cells containing a gas of Li and K atoms, respectively. \label{fig:setup-tof}}
\end{figure*}

\Autoref{fig:setup-tof} shows a schematic of the experimental setup. Its description here is brief since a detailed account was already given in Ref.~\cite{kristensen_laser-induced_2023}. A continuous beam of He droplets is produced by expanding a helium gas at a stagnation pressure of 50 bar at a temperature of 16 K into vacuum. The average radius of the droplets is $\sim$5~nm corresponding to $\sim$10$^4$ He atoms in each droplet~\cite{toennies_superfluid_2004}. The droplet beam passes through two successive pickup cells. The first (second) contains a gas of Ak (Ak$^{\prime}$) atoms. The vapor pressure in the cells is adjusted such that some of the droplets pick up two alkali atoms resulting in the formation of either an Ak$_2$, an Ak$^{\prime}_2$ or an AkAk$^{\prime}$ dimer. Afterwards, the doped He droplet beam enters a velocity map imaging (VMI) spectrometer consisting of a repeller (R), extractor (E), and a ground (G) electrode. The droplet beam is crossed by a single, focused laser beam in between the repeller and the extractor electrodes. The laser beam contains linearly polarized pulses with a duration of 50 fs (FWHM), a central wavelength of 800 nm, and an intensity of $\SI{1.3e14}{\intensity}$. The pulses are used to double ionize the alkali dimers and thereby induce Coulomb explosion. The $\Akplus$ and $\Akmplus$ fragment ions are projected onto a position sensitive detector backed by a TPX3CAM detector~\cite{fisher-levine_timepixcam_2016,nomerotski_imaging_2019,zhao_coincidence_2017} synchronized to the 1 kHz repetition rate of the laser pulses. The TPX3CAM allows us to determine the two-dimensional velocity and the time-of-flight of each ion hit. Due to the high time resolution (better than 2 ns) of the TPX3CAM, the time-of-flight measurements enable the simultaneous recording of the 2D velocity images of both the $\Akplus$ and the $\Akmplus$ ions~\cite{fisher-levine_time-resolved_2018,albrechtsen_observing_2023}. This is crucial for the implementation of the coincident filtering of the fragment ions. The 2D velocity images constitute the basic experimental observables.

\section{Results and discussion}\label{sec:results}

\subsection{Doping with Li and K atoms}\label{sec:LiK}

\subsubsection{Ion images, covariance maps and kinetic energy distributions from the mixed sample}

In the first set of measurements, the first (second) doping cell contained Li (K) vapor, as indicated on \autoref{fig:setup-tof}(a). \Autoref{fig:setup-tof}(b) shows the mass-to-charge, $m/q$, spectrum obtained from the time-of-flight recorded by the TPX3CAM detector.
The raw TPX3CAM data is a list of events representing single triggered pixels. This list of events is first time-walk corrected, then clustered and centroided, and finally time-walk corrected again, in a procedure very similar to the one described in Ref. \cite{zhao_coincidence_2017}. A publicly available custom implementation of the DBSCAN algorithm (see Data Availability) makes this post processing run approximately ten times faster than the data acquisition. The result is a list of ion hits, recording the two components of projected velocity, the time-of-flight and the corresponding lasershot number for each. The mass-to-charge is then calculated for each hit, and binned in a histogram to give the spectrum seen in \Autoref{fig:setup-tof}(b). The two peaks of interest are the ones with $m/q=7$ and $m/q=39$, assigned as \ce{^7Li+} and \ce{^39K+} ions, respectively~\footnote{\ce{^7Li+} (\ce{^39K+}) is the main isotope with an abundance of 95.2 \% (93.3 \%)}. The peak at $m/q=18$ comes from \ce{H2O+} ions created by multiphoton ionization of residual water molecules in the target chamber housing the VMI spectrometer.

\begin{figure*}
\includegraphics[width = 17.8 cm]{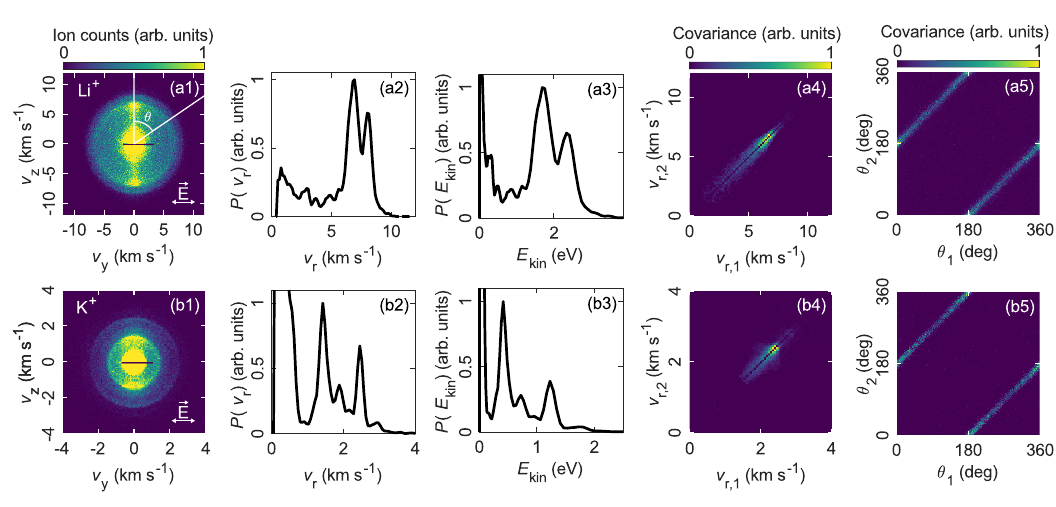}
\caption{(a1)-(a5) \ce{^7Li+} results and (b1)-(b5) \ce{^{39}K+} results.  (a1),(b1) 2D velocity images. (a2),(b2): Radial velocity distributions $P(v_r)$. (a3),(b3) Kinetic energy distributions, $P(E_{\text{kin}})$. (a4),(b4) Covariance maps of $P(v_r)$. (a5): Covariance map of the angular distribution for \ce{Li+} ions with 5.3 km/s $\leq v_r \leq$ 7.6 km/s. (b5) The same as (a5) but for \ce{K+} ions with 2.2 km/s $\leq v_r \leq$ 2.8 km/s. \label{fig:images-covariances}}
\end{figure*}

The 2D velocity image of \ce{Li+} ions, displayed in \autoref{fig:images-covariances}(a1), exhibits pronounced signal in the central region and in the outer region. The radial velocity distribution in the detector plane, $P(v_r)$, obtained by angular integration of the image, shows that the outer region consists of two overlapping peaks centered at $v_r= 6.8$~km/s and at $v_r= 8.1$~km/s, see \autoref{fig:images-covariances}(a2). Similarly, two peaks appear in the kinetic energy distribution $P(E_{\text{kin}})$ of the \ce{Li+} ions, \autoref{fig:images-covariances}(a3), obtained by first Abel inverting the 2D image using the POP algorithm~\cite{roberts_toward_2009}, and then applying a Jacobian transformation to the resulting velocity distribution~\cite{kristensen_quantum-state-sensitive_2022}. The central positions of the peaks are 1.69 eV and 2.38 eV.  The lower value matches the kinetic energy a \ce{Li+} ion acquires upon Coulomb explosion of \ce{Li2} in the 1~$^3\Sigma_{u}^+$ state, as established by recent studies on samples with only \ce{Li2}~\cite{kristensen_quantum-state-sensitive_2022}.

To confirm this interpretation, we determined the covariance map of the radial velocity distribution~\cite{christiansen_laser-induced_2016}, denoted as cov($v_r$,$v_r$) and displayed in \autoref{fig:images-covariances}(a4). The symmetric, elongated signal that peaks at at (6.8~km/s, 6.8~km/s) shows that \ce{Li+} ions with this speed or, equivalently, with $E_{\text{kin}}$ around 1.69 eV are correlated with another \ce{Li+} ion having the same speed -- or kinetic energy. Only double ionization of \ce{Li2} and subsequent Coulomb explosion of \ce{Li2^{2+}} can create such a pair of correlated \ce{Li+} ions. For completeness, we also determined the angular covariance map~\cite{hansen_control_2012,frasinski_covariance_2016,vallance_covariance-map_2021,schouder-arpc} of these \ce{Li+} ions, using the angular probability distribution in the detector plane for ions with $v_r$ between 5.3 and 7.6 km/s. The result, displayed in \autoref{fig:images-covariances}(a5) exhibits two distinct diagonal lines centered at $\theta_2 = \theta_1 \pm \SI{180}{\degree}$, where $\theta_i$, $i=1,2$ is the angle between an ion hit and the vertical center line, \autoref{fig:images-covariances}(a1). These lines demonstrate correlation between two \ce{Li+} ions recoiling with a relative angle of $\SI{180}{\degree}$. Again, such a correlation confirms that the ions originate from Coulomb explosion. Finally, we note that the signal in the central area of the image comes from dissociative ionization of dimers and possibly from \ce{Li+} ions formed by ionization of \ce{Li} atoms on droplets that picked up only a single \ce{Li} atom. Isolated \ce{Li} atoms that diffused into the VMI spectrometer will also be ionized. As this signal is rather intense it has been removed from the image, shown by the elongated rectangle.

The results for the \ce{K+} ions, displayed in \autoref{fig:images-covariances}(b1)-(b5), are similar to the \ce{Li+} results. In the 2D velocity image, \autoref{fig:images-covariances}(b1), there is a strong signal in the central region, again ascribed to atomic ionization and dissociative ionization, and then there are two main channels at larger radii. These channels are more separated than in the \ce{Li+} case, which can also be seen in $P(v_r)$, where two distinct peaks, centered at $v_r= 1.4$~km/s and at $v_r= 2.5$~km/s, stand out, \autoref{fig:images-covariances}(b2), and in $P(E_{\text{kin}})$ with one peak at 0.42 eV and another at 1.23 eV, \autoref{fig:images-covariances}(b3). The latter kinetic energy matches that of \ce{K+} ions from Coulomb explosion of \ce{K2} in the 1~$^3\Sigma_{u}^+$ state. Like the \ce{Li+} case, the symmetrically positioned island in the covariance map of the radial velocity distribution at (2.5~km/s, 2.5~km/s), \autoref{fig:images-covariances}(b4), confirms Coulomb explosion of \ce{K2} 1~$^3\Sigma_{u}^+$ dimers as the origin of these ions and so does the diagonal stripes in the corresponding angular covariance map, \autoref{fig:images-covariances}(b5).

Now we discuss the origin of the peak at $E_{\text{kin}}=2.38$~eV in $P(E_{\text{kin}})$ for \ce{Li+} and the peak at $E_{\text{kin}}=0.42$~eV in $P(E_{\text{kin}})$ for \ce{K+}. The equilibrium internuclear distance of \ce{LiK} in the 1~$^3\Sigma_{u}^+$ state is 4.97 \AA~\cite{rousseau_theoretical_1999}. If this heteronuclear dimer is doubly ionized, the resulting \ce{LiK^{2+}} dication is created with a potential energy of 2.90 eV, in the Coulomb potential approximation, compared to a \ce{Li+}, \ce{K+} ion pair at infinite distance. Upon Coulomb explosion, the \ce{Li+} ion ends with $E_{\text{kin}}=2.46$~eV and the \ce{K+} ion with $E_{\text{kin}}=0.44$~eV due to conservation of momentum. These two values are very close to the center values of the two above mentioned peaks in $P(E_{\text{kin}})$ for \ce{Li+} and \ce{K+}, respectively. Thus, we interpret the 2.38~eV peak in $P(E_{\text{kin}})$ for \ce{Li+} and the 0.42~eV peak in $P(E_{\text{kin}})$ for \ce{K+} as originating from laser-induced Coulomb explosion of \ce{LiK} in the 1~$^3\Sigma^+$ state into a (\ce{Li+},\ce{K+}) pair.

\begin{figure}
\includegraphics[width = 8.6 cm]{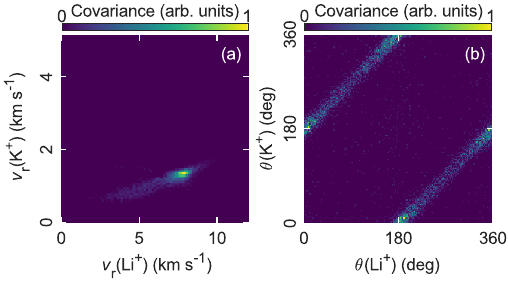}
\caption{(a) Covariance map of the radial velocity distribution between the \ce{Li+} ions and the \ce{K+} ions. (b) Covariance map of the angular distribution between the \ce{Li+} ions with 7.5 km/s $\leq v_r \leq$ 9.6 km/s and the \ce{K+} ions with 1.0 km/s $\leq v_r \leq$ 1.8 km/s.   \label{fig:hetero-covariance}}
\end{figure}

To corroborate this interpretation, we determine the covariance map of the radial velocity distribution for the \ce{Li+} and the \ce{K+} ions, see \autoref{fig:hetero-covariance}(a). The strong covariance signal centered at (8.1~km/s, 1.4~km/s) shows that a \ce{Li+} with $v_r$ around 8.1~km/s is correlated with a \ce{K+} ion with $v_r$ around 1.4~km/s. This can also be formulated as when a \ce{Li+} ion with $E_{\text{kin}}=2.38$~eV is detected, it is likely to also detect a \ce{K+} ion wih $E_{\text{kin}}=0.42$~eV. The only way to produce such a correlated (\ce{Li+}, \ce{K+}) ion pair is through double ionization of \ce{LiK} in the 1~$^3\Sigma^+$ state and subsequent Coulomb explosion of the \ce{LiK^{2+}} dication. As for the homonuclear dimers, we determined the covariance map between the angular distribution of the \ce{Li+} ions pertaining to the 2.38~eV peak and the \ce{K+} ions in the 0.42~eV peak. The covariance map, displayed in \autoref{fig:hetero-covariance}(b), shows correlation at a relative emission angle of $\SI{180}{\degree}$ as expected for Coulomb explosion of \ce{LiK}.

\subsubsection{Coincident filtering: Separation of fragment ions from homonuclear and heteronuclear dimers}

So far we have shown that laser-induced Coulomb explosion reveals the presence of both homonuclear and heteronuclear dimers in our sample. Now we discuss how coincident filtering allows the separation of the \ce{Li+} (\ce{K+}) fragment ions that originate from either \ce{Li2} (\ce{K2}) or from \ce{LiK} parent dimers. The filter relies on the exact division of momentum amongst the two fragment ions in the Coulomb explosion of a dimer. Thus, the filter for \ce{Li2} (\ce{K2}) works by determining those \ce{Li+} ions that, within the same laser shot, are detected along with another \ce{Li+} (\ce{K+}) ion having the diametrical momentum vector. Such ions can only be produced by Coulomb explosion of \ce{Li2} (\ce{K2}), i.e., the filter discards \ce{Li+} (\ce{K+}) produced from Coulomb explosion of either \ce{LiK} dimers or of \ce{Li3} (\ce{K3}), \ce{Li2K}, \ce{LiK2} trimers~\footnote{Very few trimers are actually formed because the alkali vapor pressures in the pickup cells are kept low}, and from dissociative ionization of either \ce{Li2} (\ce{K2}) or of LiK. Likewise, the filter for \ce{LiK} acts by determining the \ce{Li+} ions that, within the same laser shot, are detected along with a \ce{K+} ion having the opposite momentum.

\begin{figure}
\includegraphics[width = 8.6 cm]{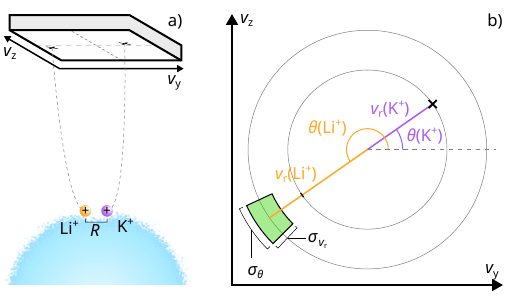}
\caption{(a) Illustration of the detection of the 2D velocity ($v_y$, $v_z$) of the \ce{Li+} and \ce{K+} fragment ions from Coulomb explosion of \ce{LiK}. (b) Illustration of the principle of the coincidence filter for the \ce{LiK} dimer. Here $\sigma_{v_r}$ and $\sigma_{\theta}$ define the velocity and angular tolerance of the filter. The values used are given in \autoref{eq:LiK-Li-filter} and \autoref{eq:LiK-K-filter}. \label{coincidence-filter-schematic}}
\end{figure}

In practice, the filter is implemented in a polar coordinate system covering the detector plane, see \autoref{coincidence-filter-schematic}. Two \ce{Li+} ions are considered coincident if the following conditions are fulfilled
\begin{equation}
\label{eq:Li2-filter}
\begin{aligned}
\mid \theta_1(\ce{Li+})-\theta_2(\ce{Li+})-\SI{180}{\degree}\mid <  \SI{15}{\degree}\\
\mid v_{r,1}(\ce{Li+})-v_{r,2}(\ce{Li+})\mid <  \text{1.3}~\text{km/s},
\end{aligned}
\end{equation}
where lower index 1 and 2 refer to two different \ce{Li+} ions. \Autoref{eq:Li2-filter} defines the filter for identifying a \ce{Li2} homonuclear dimer. The tolerances of the filter are chosen to encompass the full covariance lines seen in \autoref{fig:images-covariances}(a4) and (a5). Similarly, the filter for identifying a \ce{K2} homonuclear dimer is defined by:
\begin{equation}
\label{eq:K2-filter}
\begin{aligned}
\mid \theta_1(\ce{K+})-\theta_2(\ce{K+})-\SI{180}{\degree}\mid <  \SI{15}{\degree}\\
\mid v_{r,1}(\ce{K+})-v_{r,2}(\ce{K+})\mid <  \text{0.6}~\text{km/s},
\end{aligned}
\end{equation}
and the filter for identifying the \ce{LiK} heteronuclear dimer:
\begin{equation}
\label{eq:LiK-Li-filter}
\begin{aligned}
\mid \theta(\ce{K+})-\theta(\ce{Li+})-\SI{180}{\degree}\mid <  \SI{20}{\degree}\\
\mid \frac{m_\text{K}}{m_\text{{Li}}}v_{r}(\ce{K+})-v_{r}(\ce{Li+})\mid <  \text{1.3}~\text{km/s},
\end{aligned}
\end{equation}
when detecting \ce{Li} ions, and,
\begin{equation}
\label{eq:LiK-K-filter}
\begin{aligned}
\mid \theta(\ce{K+})-\theta(\ce{Li+})-\SI{180}{\degree}\mid <  \SI{20}{\degree}\\
\mid v_{r}(\ce{K+})-\frac{m_\text{{Li}}}{m_\text{{K}}}v_{r}(\ce{Li+})\mid <  \text{0.6}~\text{km/s},
\end{aligned}
\end{equation}
when detecting \ce{K+} ions. Here, $m_\text{K} = 39$~u and $m_\text{Li} = 7$~u. The mass ratios are needed, as the even momentum sharing results in two ions with different speeds.

\begin{figure}
\includegraphics[width = 8.6 cm]{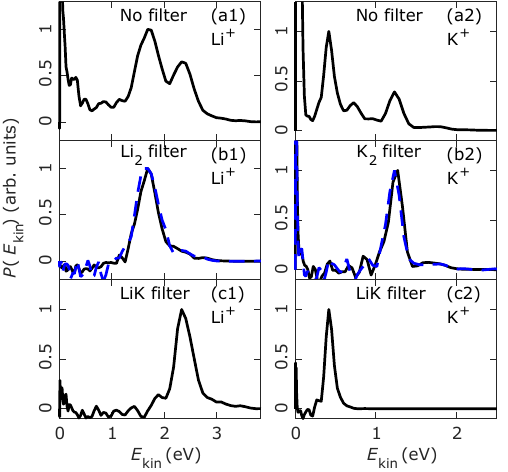}
\caption{Full black curves: Kinetic energy distributions of \ce{Li+} ions, panels (a1)-- (c1), and of \ce{K+} ions, panels (a2)--(c2), determined from those ions that passed the coincidence filter indicated on the top of each panel. Blue dashed lines: Kinetic energy distributions of \ce{Li+} ions, panel (b1), and of \ce{K+} ions, panels (b2) obtained from droplets doped only with Li or K. \label{fig:filtered-Ekin}}
\end{figure}

Now we apply the \ce{Li2} coincidence filter, \autoref{eq:Li2-filter}, to the \ce{Li+} ions detected and determine $P(E_{\text{kin}})$ from those ions that passed the filter. The result is shown by the full black curve in \autoref{fig:filtered-Ekin}(b1). As a reference, we plot $P(E_{\text{kin}})$ for \ce{Li+} ions, blue dashed curve, recorded with an empty potassium pickup cell, i.e., the helium droplets only picked up \ce{Li} atoms and thus the only dimers formed were \ce{Li2}. The good agreement between the two curves demonstrates the desired effect of the filter to select the fragment ions corresponding to a particular dimer species, in this case \ce{Li2}, from the mixed sample.  Note that $P(E_{\text{kin}})$ obtained via the coincidence filter exhibits only a single peak corresponding to dimers in the 1~$^3\Sigma_{u}^+$ state, i.e., there is no evidence of dimers formed in the 1~$^1\Sigma_{g}^+$ state, an observation consistent with previous experiments on samples of pure \ce{Li2} dimers~\cite{lackner_spectroscopy_2013,kristensen_quantum-state-sensitive_2022,kristensen_laser-induced_2023}.

In a similar manner, we apply the \ce{K2} coincidence filter, \autoref{eq:K2-filter}, to the \ce{K+} ions detected and determine $P(E_{\text{kin}})$ from the ions that passed this filter. The result is shown by the full black curve in \autoref{fig:filtered-Ekin}(b2). Again, we include a reference result, blue dashed curve, depicting $P(E_{\text{kin}})$ for \ce{K+} ions originating from helium droplets doped only with potassium. The good agreement between the two curves shows that we can select the \ce{K+} ions pertaining to the \ce{K2} dimers from the mixed sample. Note that in addition to the main peak, centered at 1.23 eV, corresponding to dimers in the 1~$^3\Sigma_{u}^+$ state, there is a smaller peak centered at 1.70 eV. The latter is assigned to \ce{K2} dimers in the 1~$^1\Sigma_{g}^+$ state. This is also consistent with previous experiments on samples of pure \ce{K2} dimers~\cite{kristensen_quantum-state-sensitive_2022,kristensen_laser-induced_2023}.

Having established that the coincidence filtering method works well for retrieving $P(E_{\text{kin}})$ for Coulomb explosion of both \ce{Li2} and \ce{K2} in the mixed sample, we assume that it will also work well for retrieving the signal from the heteronuclear dimer. Thus, we apply the \ce{LiK} coincidence filter to both the \ce{Li+} ions, \autoref{eq:LiK-Li-filter}, and to the \ce{K+} ions, \autoref{eq:LiK-K-filter}, and determine $P(E_{\text{kin}})$ in both cases. The results are displayed in \autoref{fig:filtered-Ekin}(c1) and (c2). The curves reach their maxima at 2.38 eV and 0.42 eV, respectively, very close to the values 2.46 eV and 0.44 eV expected for fragmentation of the \ce{LiK^{2+}} dication via a Coulomb potential starting from the equilibrium distance of dimers in the 1~$^3\Sigma^+$  state. Both kinetic energy distributions display only a single peak indicating that the formation of \ce{LiK} dimers in the singlet state is insignificant under our current experimental conditions.

\subsubsection{Distribution of internuclear distances for \ce{LiK}}

In the former studies on Coulomb explosion of the homonuclear alkali dimers, the distribution of internuclear distances $P(R)$ was determined from $P(E_{\text{kin}})$  via the ($R$, $E_\text{kin}$) relation using either the Coulomb potential for the dications or a more accurate potential from quantum chemistry calculations. For \ce{Li2}, \ce{K2} and \ce{Rb2} in the 1~$^3\Sigma_{u}^+$ state, it was found that the center of the measured $P(R)$ was within 0.05~\AA~of the center of the square of the wave function $|\Psi(R)|^2$ for the vibrational ground state, whereas the FWHM of $P(R)$ was about a factor of 2 larger than that of $|\Psi(R)|^2$~\cite{kristensen_laser-induced_2023}.

\begin{figure}
\includegraphics[width = 8.6 cm]{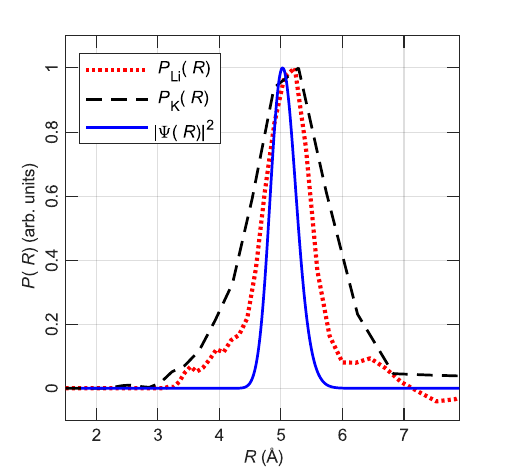}
\caption{The distribution of internuclear distances for \ce{LiK} determined from the \ce{K+} fragment ions (dashed black curve) and from the \ce{Li+} fragment ions (dotted red curve). The full blue curve shows the calculated square of the internuclear wave function for the vibrational ground state of \ce{LiK} in the 1~$^3\Sigma^+$ state. \label{fig:LiK-R-distributions}}
\end{figure}

In the case of \ce{LiK}, it should be possible to determine $P(R)$ from either the \ce{Li+} fragment ions or from the \ce{K+} fragment ions. In \autoref{fig:LiK-R-distributions}, the red dotted curve and the black dashed curve show $P(R)$ obtained from $P(E_{\text{kin}})$ for the \ce{Li+} ions and from the \ce{K+} ions, respectively. As a reference we also plot $|\Psi(R)|^2$, the blue full curve obtained by solving the stationary vibrational Schr\"{o}dinger equation with the internuclear potential from the literature~\cite{rousseau_theoretical_1999}. First, we note that $P_\text{Li}(R)$ and $P_\text{K}(R)$, where the lower index labels which ion fragment the distribution is obtained from, have almost the same center positions, 5.10~\AA~and 5.14~\AA, i.e. values that are very close to that of the center of $|\Psi(R)|^2$, 5.04 \AA. Second, we note that the full width at half maximum (FWHM) of $P_\text{K}(R)$, 1.5~\AA, is significantly larger than the FWHM of $P_\text{Li}(R)$, 0.86~\AA.

If the kinetic energy distributions measured for the two fragments, $P_\text{K}(E_{\text{kin}})$ and $P_\text{Li}(E_{\text{kin}})$, were solely determined by $|\Psi(R)|^2$, then $P_\text{K}(R)$ and $P_\text{Li}(R)$ would be identical. This is, however, not the case. As discussed in Ref.~\cite{kristensen_laser-induced_2023}, the kinetic energy distributions are broadened by various experimental factors such as the energy resolution of the VMI spectrometer and nuclear motion in the double ionization process. Since the \ce{K+} and \ce{Li+} ions share the momentum equally in the Coulomb explosion, the \ce{K+} fragment acquires only 7/46 of the total Coulomb energy whereas the \ce{Li+} fragment acquires a fraction of 39/46. Thus, in the transformation of the initial probability distribution $P(E_{\text{kin}})$ to the final probability distribution $P(R)$ any broadening of $P(E_{\text{kin}})$ will be amplified much more (by a factor of 39/7) for the \ce{K+} case as compared to the \ce{Li+} case. This explains why $P_\text{K}(R)$ is broader than $P_\text{Li}(R)$. For comparison, the FWHM of $|\Psi(R)|^2$ is 0.50~\AA, i.e., the FWHM of $P_\text{Li}(R)$ is about 70\% larger than this value.

\subsection{Doping with \ce{Na} and \ce{K} atoms}\label{sec:NaK}

To test our method for a case where the peaks in the kinetic energy spectrum of the fragment ions from the homonuclear and heteronuclear dimers overlap more strongly than in the Li-K case, we carried out measurements on a mixed sample of \ce{Na2}, \ce{K2} and \ce{NaK}. This was obtained by having a Na vapor in the first and a K vapor in the second pickup cell.

\begin{figure}
\includegraphics[width = 8.6 cm]{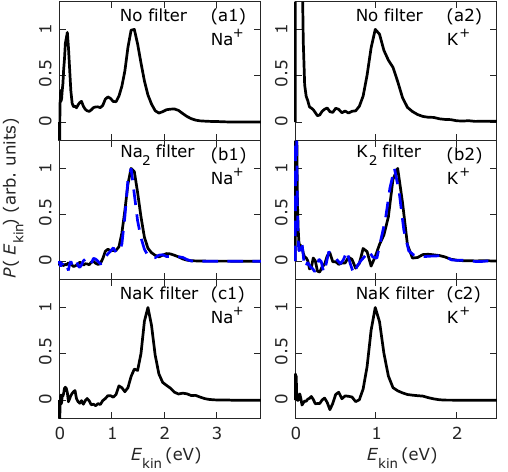}
\caption{Full black curves: Kinetic energy distributions of \ce{Li+} ions, panels (a1), (b1), and of \ce{K+} ions, panels (a2), (b2), determined from those ions that passed the coincidence filter indicated on the top of each panel. Blue dashed lines: Kinetic energy distributions of \ce{Li+} ions, panel (a1), and of \ce{K+} ions, panels (a2) obtained from droplets doped only with Li. \label{fig:NaK-Ekin}}
\end{figure}

\Autoref{fig:NaK-Ekin}(a1) shows $P(E_{\text{kin}})$ for the \ce{Na+} ions obtained directly from the 2D image, i.e., without any coincident filtering. The 2D velocity images, radial velocity distributions and covariance maps, similar to those for the Li-K case displayed in \autoref{fig:images-covariances} and in \autoref{fig:hetero-covariance}, are given in the appendix. The kinetic energy spectrum is dominated by a large peak centered at 1.41~eV and a small peak centered at 2.13 eV. Unlike the Li-K case, the major peak at 1.41~eV is not split in two. However, upon application of first the \ce{Na2} filter, full black curve in \autoref{fig:NaK-Ekin}(b1), and then the \ce{NaK} filter, \autoref{fig:NaK-Ekin}(c1), it appears that the 1.41~eV peak is indeed made up of two contributions. In analogy with the analysis of the Li-K data, the full black curve in \autoref{fig:NaK-Ekin}(b1) should represent $P(E_{\text{kin}})$ for \ce{Na2}. The good agreement with the dashed blue curve, showing $P(E_{\text{kin}})$ for \ce{Na+} ions from droplets that only picked up \ce{Na} atoms (empty \ce{K} pickup cell) supports the accuracy and intended effect of the filter. The curve in \autoref{fig:NaK-Ekin}(c1) should then be $P(E_{\text{kin}})$ for \ce{NaK}. This assignment is consistent with the fact that the peak position, 1.68~eV is very close to the value, 1.65~eV, a \ce{Na+} ion acquires upon Coulomb explosion of \ce{NaK} in the 1~$^3\Sigma^+$ state. The tiny peak at $\sim$~2.5 eV indicates that a small fraction of the NaK dimers are formed in the 1~$^1\Sigma^+$ state since \ce{Na+} ions from Coulomb explosion of dimers in this state will get a kinetic energy of 2.6 eV~\cite{deiglmayr_calculations_2008}.

\Autoref{fig:NaK-Ekin}(a2) shows $P(E_{\text{kin}})$ for the \ce{K+} ions without any coincident filtering. A single borad peak is observed with an indication of a double peak structure. Upon application of first the \ce{K2} filter, full black curve in \autoref{fig:NaK-Ekin}(b2), and then the \ce{NaK} filter, \autoref{fig:NaK-Ekin}(c2) to the \ce{K+} ions, the two individual peaks are revealed. Again, we plot $P(E_{\text{kin}})$ obtained using droplets that picked up only \ce{K} atoms, dashed blue curve in \autoref{fig:NaK-Ekin}(b2). As for the \ce{Na+} ions, there is good agreement with $P(E_{\text{kin}})$ extracted from the mixed sample. The majority of the \ce{K2} dimers are created in the 1~$^3\Sigma_{u}^+$ state (peak at 1.25~eV) although \ce{K2} in the 1~$^1\Sigma_{g}^+$ state are also present (peak at 1.7~eV). Finally, the peak position of $P(E_{\text{kin}})$ obtained with the \ce{NaK} filter is 1.01~eV, which is close to the value expected for a \ce{K+} ion created when \ce{NaK} at the equilibrium distance of the 1~$^3\Sigma^+$ state is Coulomb exploded. Overall, the data presented in \autoref{fig:NaK-Ekin} show that for laser-induced Coulomb explosion of the mixed sample \ce{Na2}, \ce{K2} and \ce{NaK} our coincident filtering method is capable of extracting the individual $P(E_{\text{kin}})$ pertaining to these three dimer species, unaffected by the strongly overlapping kinetic energy channels.

\begin{figure}
\includegraphics[width = 8.6 cm]{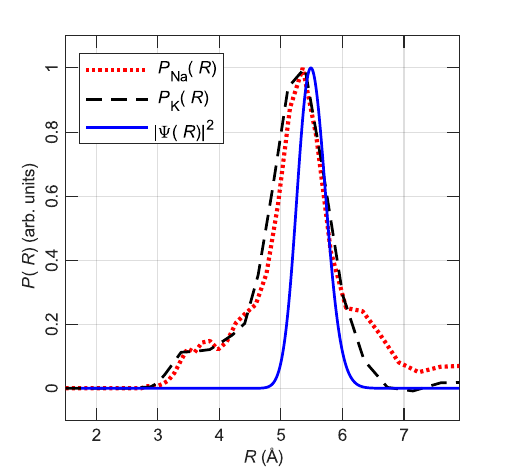}
\caption{The distribution of internuclear distances for \ce{NaK} determined from the \ce{K+} fragment ions (dashed black curve) and from the \ce{Na+} fragment ions (dotted red curve). The full blue curve shows the calculated square of the internuclear wave function for the vibrational ground state of \ce{NaK} in the 1~$^3\Sigma^+$ state. \label{fig:NaK-R-distributions}}
\end{figure}

At last, we also determine $P(R)$ of \ce{NaK} using either $P(E_{\text{kin}})$ for the \ce{Na+} ions or for the \ce{K+} ions. \Autoref{fig:NaK-R-distributions} shows $P_\text{Na}(R)$, red dotted curve, and $P_\text{K}(R)$, black dashed curve and, as a reference, $|\Psi(R)|^2$ for the vibrational ground state of the 1~$^3\Sigma^+$ state, full blue curve, obtained by solving the stationary vibrational Schr\"{o}dinger equation using the internuclear potential~\cite{aymar_calculations_2007}. The center positions of $P_\text{Na}(R)$ and of $P_\text{K}(R)$ are 5.33~\AA and 5.28~\AA, which are close to the center of $|\Psi(R)|^2$, 5.50~\AA. The FWHM of $P_\text{Na}(R)$ is 0.92~\AA{} and 1.1~\AA{} for $P_\text{K}(R)$. This similarity is expected since the smaller mass difference between \ce{Na} and \ce{K} leads to a ratio of only 39/23 for $E_{\text{kin}}$ of the \ce{Na+} and the \ce{K+} ions compared to an $E_{\text{kin}}$-ratio between \ce{Li+} and \ce{K+} of 39/7 in the Li-K case. The FWHM of $|\Psi(R)|^2$ is 0.55~\AA, i.e., the experimental $P(R)$ distributions are about a factor of two broader. This is similar to the case of the homonuclear dimers previously studied~\cite{kristensen_laser-induced_2023}.

\section{Conclusion and outlook}\label{sec:conclusion-outlook}

In this work, we have demonstrated a method on how to separate the different contributions to one fragment ion species stemming from laser-induced Coulomb explosion of different parent molecules present in a mixed sample. Our work concerned a heteronuclear alkali dimer, like \ce{LiK}, created on the surface of a helium nanodroplet in which case the target sample also contained droplets doped with one of the two homonuclear alkali dimers, i.e., \ce{Li2} and \ce{K2}. We showed that the \ce{Li+} and \ce{K+} fragment ions originating from either \ce{Li2}, \ce{K2}, or \ce{LiK} can be individually filtered out using a coincidence method based on the momentum conservation in the Coulomb explosion process. This enabled us to determine the kinetic energy distribution of the \ce{Li+} and \ce{K+} ions from Coulomb explosion of \ce{LiK} and in turn the distribution of the internuclear distances of the initial heteronuclear dimer. To demonstrate the versatility of the coincidence filter technique, we also showed that it works as desired for \ce{NaK} in a mixed sample of \ce{NaK}, \ce{Na2}, and \ce{K2}.

Our method should apply generally to any mixed samples of homonuclear and heteronuclear dimers of alkali atoms or of alkali-alkaline earth dimers~\cite{lackner_helium-droplet-assisted_2014,lackner_photoinduced_2018}. This also includes the separation of naturally occurring isotopologues for a dimer containing only one alkali species like \ce{^7Li^6Li}, \ce{^6Li2}, and \ce{^7Li2}. Furthermore, the technique could be principally extended to mixed alkali trimers and higher order oligomers.  An immediate use of the coincident filtering technique is in time-resolved measurement of rotational and vibrational motion of dimers (or trimers)  since the different masses of the constituents lead to different nuclear dynamics. Furthermore, the filter technique should also apply to identify droplets doped with two specific dopants inside a droplet~\cite{choi_infrared_2006,yang_helium_2012,pickering_femtosecond_2018,schouder_laser-induced_2021} or with an alkali atom on the surface and a molecule in the interior~\cite{renzler_communication:_2016,albrechtsen_observing_2023}. Such identification could open possibilities for exploring ion-molecule complex formation when the alkali atom is selectively ionized by a femtosecond laser pulse.

\section*{DATA AVAILABILITLY}
The source data presented in this work and the analysis scripts used to treat them can be obtained from the authors on reasonable request. The optimized clustering and centroiding program for the TPX3CAM data is publicly available online at: \url{https://github.com/laulonskov98/pixel_centroiding_DBSCAN}.

\begin{acknowledgments}
We thank Jan Th{\o}gersen for expert help on keeping the laser system in optimal condition. We are grateful to Laurits Lønskov Sørensen and Simon Fischer-Nielsen for their great help in writing the optimized program for clustering and centroiding the TPX3CAM data. H.S. acknowledges support from Villum Fonden through a Villum Investigator Grant No. 25886.
\end{acknowledgments}

\vspace{2 cm}

\appendix*\section{NaK: 2D velocity images, radial velocity distributions and covariance maps}
\label{sec:Appendix}

\begin{figure*}
\includegraphics[width = 17.8 cm]{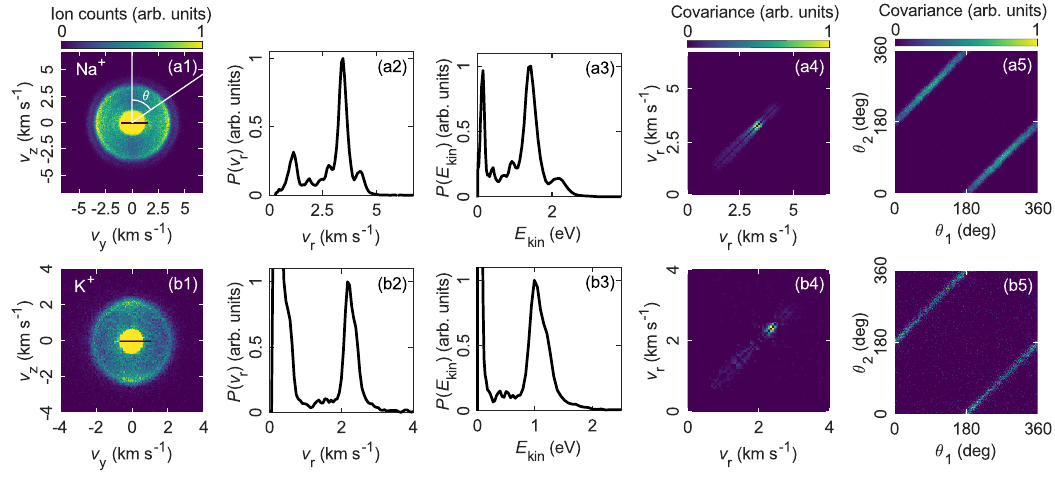}
\caption{(a1)-(a5) \ce{^23Na+} results and (b1)-(b5) \ce{^{39}K+} results.  (a1),(b1) 2D velocity images. (a2),(b2) Radial velocity distributions $P(v_r)$. (a3),(b3) Kinetic energy distributions $P(E_{\text{kin}})$. (a4),(b4) Covariance maps of $P(v_r)$. (a5) Covariance map of the angular distribution for \ce{Na+} ions with 2.9 km/s $\leq v_r \leq$ 4.0 km/s. (b5) The same as (a5) but for \ce{K+} ions with 1.1 km/s $\leq v_r \leq$ 3.3 km/s.  \label{fig:app:images-covariances}}
\end{figure*}

\begin{figure}
\includegraphics[width = 8.6 cm]{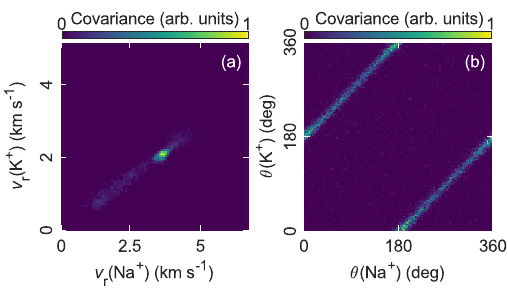}
\caption{(a) Covariance map of the radial velocity distribution between the \ce{Na+} ions and the \ce{K+} ions. (b) Covariance map of the angular distribution between the \ce{Na+} ions with 2.9 km/s $\leq v_r \leq$ 4.3 km/s and the \ce{K+} ions with 1.6 km/s $\leq v_r \leq$ 2.5 km/s. \label{fig:app:hetero-covariance}}
\end{figure}

\Autoref{fig:app:images-covariances} and \Autoref{fig:app:hetero-covariance} are similar to \autoref{fig:images-covariances} and \autoref{fig:hetero-covariance} but show the experimental results obtained when the helium droplets pass through the doping cells containing Na vapor in the first and K vapor in the second pickup cell. In particular, \autoref{fig:app:hetero-covariance} displays the radial velocity map and the angular covariance map of the \ce{Na+} and \ce{K+} fragment ions. The clear correlation corroborates that the ions originate from Coulomb explosion of \ce{NaK}, and thereby validate the filtering used to extract the results shown in Sec. \ref{sec:NaK}.

\newpage

%



\end{document}